\documentstyle[preprint,eqsecnum,aps,tighten]{revtex}
\input{psfig}

\def\D0{\mbox{D\O }}
\def\Et{$E_{T}$}

\def\MEt{\mbox{$E\kern-0.57em\raise0.19ex\hbox{/}_{T}$}\ }
\def\MEtp{\mbox{$E\kern-0.57em\raise0.19ex\hbox{/}_{T}$}}

\begin{document}
\title{Search for Leptoquarks at \D0 
\protect\footnote{Submitted to the proceedings
of {\em Hadron Collider Physics XII}, June 4-11, 1997, 
Stony Brook, NY.}}

%
\author{Douglas M. Norman \footnote{representing the \D0 Collaboration}}
\address{Texas A\& M University, Department of Physics, College Station, TX
77843}

\maketitle

\begin{abstract} We present in this paper the current status of searches
for leptoquarks at \D0. These results include the use of next leading order
theoretical predictions for the cross section for pair production of
leptoquarks at hadron colliders. We also present a new optimized
analysis for first generation leptoquarks with significant increase in
sensitivity relative to earlier searches using \D0 data. The mass limits 
derived from this first
generation leptoquark search are relevant to the recently reported
high mass events at HERA.
\end{abstract}

\section{Introduction}

Leptoquarks are hypothetical exotic particles with both lepton and
color quantum numbers; they are color triplet bosons with fractional charge.
They are produced in pairs at hadron colliders
via strong interactions. A leptoquark will decay via an unknown coupling,
$\lambda$,
to a lepton and a quark. The production of leptoquark pairs in $p\bar{p}$
collisions
 is insensitive to
$\lambda$, and we are not concerned with this coupling as long
as it is greater than 10$^{-12}$; otherwise, the leptoquarks will not decay
in our detector and escape detection.
We assume that leptoquarks occur in  generations;
this is to say that leptoquarks of one generation couple exclusively to
leptons and quarks of the same generation. For example, a first generation
leptoquark will only couple to electrons, electron neutrinos and u or d quarks.
With this assumption of generations, we are able to bypass strict bounds 
($\ge$ 1 TeV) on leptoquark masses derived from limits on flavor changing 
neutral currents.

Since leptoquarks are produced dominantly in pairs at hadron colliders, 
their signature
would be two leptons plus two or more jets. A free parameter of the
model for leptoquarks is the branching fraction of the leptoquark to
charged lepton plus quark, $\beta$. For $\beta = 1.0$, leptoquarks 
decay 100\% to
a charged lepton plus quark. In the case for a first generation leptoquark
pair, the signature would be two isolated electrons ($e^+e^-$) plus 
at least two jets.
Additional jets could arise from initial state radiation or final
state radiation. Since we assume leptoquarks do not couple outside their
own generation, for first generation leptoquarks, we would not expect
to see muons or evidence of taus in the signature.

Some published and previous preliminary conference limits on leptoquarks masses
are shown in tables~\ref{tab:one} and \ref{tab:two}. 
These give a brief history of mass limits
from the Tevatron. 
The previously published limits from CDF and \D0 have used Leading Order (LO) 
theory. Next Leading Order (NLO) theory predicts cross sections that are 
30-50\% higher than LO theory. NLO was used for the CDF limits at recent
conferences~\cite{Moriond}\cite{vanderbilt} and will be used in the newest
limits presented in this paper from \D0.

\section{First Generation Leptoquarks}

\D0 searches for first generation leptoquarks primarily in the dielectron 
plus two jet
and electron plus missing transverse energy (\MEt) plus two jet signatures
corresponding to the decay modes $eeqq$ and $e\nu qq$.
The search for first generation leptoquarks in the dieletron plus two jet
signature 
has been greatly modified since the 
1997 Moriond Conferences~\cite{Moriond}. After the
announcement of the high mass events from HERA~\cite{hera},
\D0 has retooled the analysis to increase the acceptance for
leptoquarks by increasing the acceptance for electrons and
by optimizing the event section for higher mass leptoquarks.
The data sample used in this analysis is the full Run I
data sample of 123 pb$^{-1}$. The single electron plus \MEt plus two jets
signature search is also optimized at higher masses. 
We are also using the newly available NLO theory
predictions~\cite{nlo_theory} for the production cross section.

The electron identification for one of the two electrons in the
dielectron search has been loosened to increase the acceptance for leptoquarks.
For the loose selection, the track requirement is removed.
This has effectively doubled
the acceptance for dielectrons. The relaxing of the track requirement
for the one electron does not significantly increase the QCD background 
relative to
the tight-tight electron selection.

The basic $eejj$ event selection is given in the following list:
\begin{itemize}

\item two electrons with \Et\ $>$ 20 GeV with $\mid \eta \mid < 1.1$ or
$1.5 < \mid \eta \mid < 2.5$,

\item two jets with \Et\ $>$ 15 GeV with $\mid \eta \mid < 2.5$ where the
jets are reconstructed with a cone algorithm using a cone of radius
$R = \sqrt{\Delta\eta^2 + \Delta\phi^2} = 0.7$,

\item exclude events with dielectron invariant masses between 82 to 100
GeV/c$^2$.

\end{itemize}
The dielectron mass cut is a three sigma cut. With this basic selection,
we have 101 events in our candidate sample. The main backgrounds come from
Drell-Yan, QCD with jets faking electrons, and Top. We estimate 66.8$\pm$13.4 
events for Drell-Yan, 24.3$\pm$3.6 events from QCD, and 1.8$\pm$0.7 events
from Top for a total of 92.8$\pm$13.8 background events. 

Further event selection is  optimized for leptoquark masses of 200 GeV/c$^2$ 
and greater. The optimization is based on signal and 
background Monte Carlo
event samples and on QCD data event samples. Several variables have been 
studied
for the optimization; variables involving sums of energies and transverse 
energies (\Et) of electrons and jets, variables of event shape, variables
of reconstructed invariant masses, and variables involving constraints with 
leptoquark
masses have been used. Systematic grid search~\cite{amos} and neural 
network~\cite{nets}
techniques are used for the optimization studies. We approached the
optimization in two ways. The first is a discovery search where we optimized
signal ($S$) over the square root of the background ($B$), $S/\sqrt{B}$.
The second is for setting a limit where we maximized the signal for
a given target background of about 0.4 events. This level of background
gives a probability of about 70\% of seeing 0 background events.

\begin{figure}
\centerline{\psfig{figure=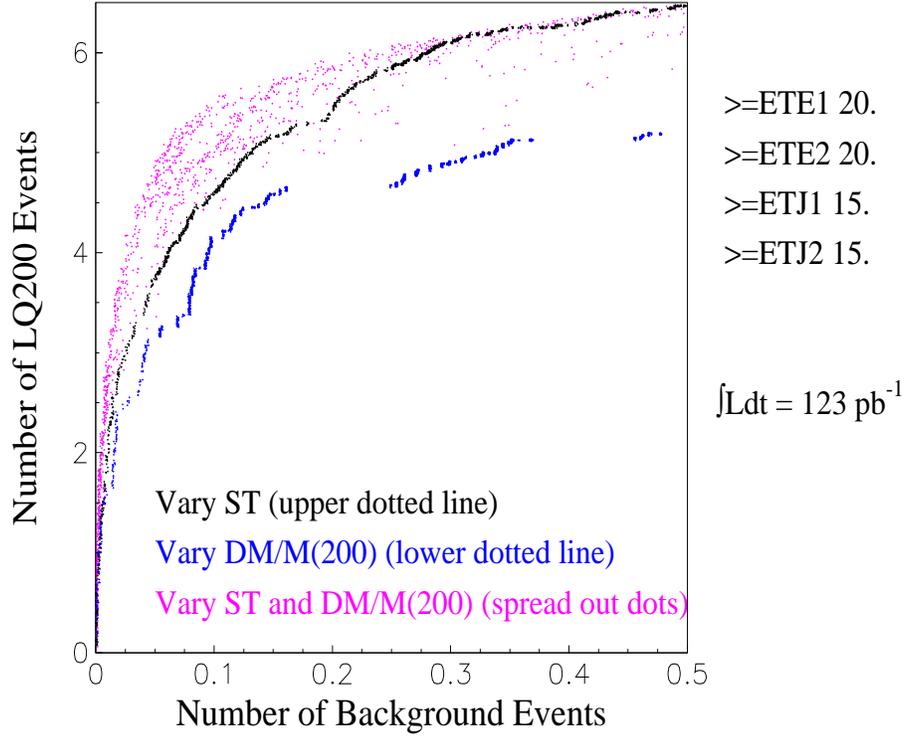,height=5.5in,width=5.0in}}
\caption{Output of grid search. We can see by this that the $S_T$ variable
alone
optimizes the signal to background ratio better than the mass constrained
variable, DM/M(200), and the combination of both.}
\label{fig:one}
\end{figure}

\begin{figure}
\centerline{\psfig{figure=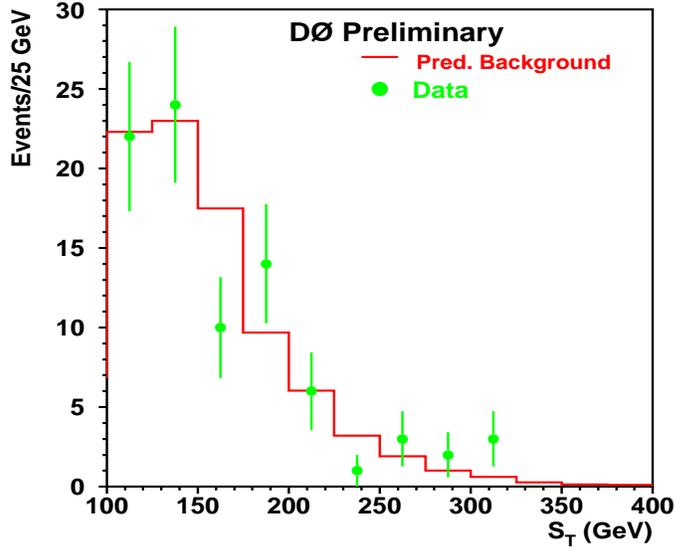,height=3.5in,width=4.0in}}
\caption{The differential distribution of $S_T$ for data and background
predictions. The basic event selection is applied.}
\label{fig:st1}
\end{figure}

\begin{figure}
\centerline{\psfig{figure=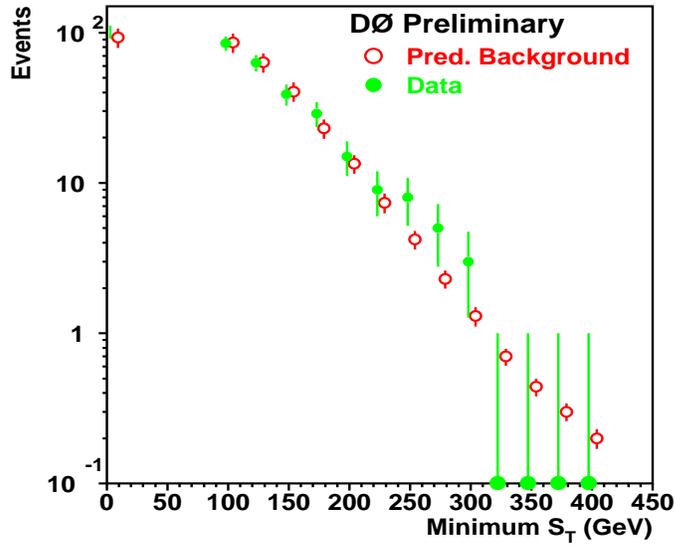,height=3.5in,width=4.0in}}
\caption{The integrated distribution of $S_T$ for data and background
predictions. The basic event selection is applied.}
\label{fig:st2}
\end{figure}

We find that a 
transverse energy
variable defined as $S_T = {\displaystyle\sum_{\mbox{\scriptsize jets}}{E_T^j}} 
+ E_T^{e1} + E_T^{e2}$
is the most optimal variable. Here, the jets are required to have \Et\ $>$ 
15 GeV to be included in the sum. Figure~\ref{fig:one} shows the output
of an example grid search. Here is plotted the number of signal events verses
the number of background events. Each point in the plot represents a 
different set of cuts. The electron and jet \Et's are fixed at the
basic cut level as given in the figure. The upper line of dots are for
the $S_T$ cut. The lower line of dots is for a leptoquark mass constrained
variable defined as: DM/M(200) = $\sqrt{(M_{ej1} - M_{LQ}^{200})^2 +
(M_{ej2} - M_{LQ}^{200})^2}/M_{LQ}^{200}$. $M_{ej1}$ and $M_{ej2}$ are
the invariant masses of the two combinations of the electrons and jets
such that the difference in the two masses is minimized, and $M_{LQ}^{200}$ is
the mass of a 200 GeV/c$^2$ mass leptoquark in this example. 
The wide band of 
dots is the combination of both variables used in the selection. From this
plot we see that that the $S_T$ variable does better than the mass constraint
variable at our target background of 0.4 events.
We find that a $S_T > 350$ GeV gives a background of about 0.4 events.
None of the 101 candidate events pass this cut. The differential
and integrated $S_T$ distributions for the data and background predictions
are given in Figs.~\ref{fig:st1} and ~\ref{fig:st2}. The background
distribution matches well the data distribution.

\begin{figure}
\centerline{\psfig{figure=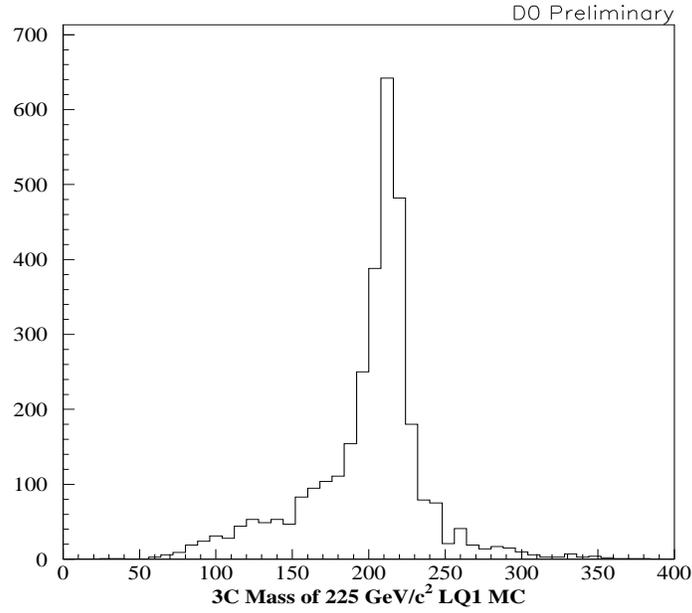,height=3.5in,width=4.0in}}
\caption{The 3C mass fit for a 225 GeV/c$^2$ mass leptoquark sample,
$\beta = 1.0$.}
\label{fig:3cmc}
\end{figure}

\begin{figure}
\centerline{\psfig{figure=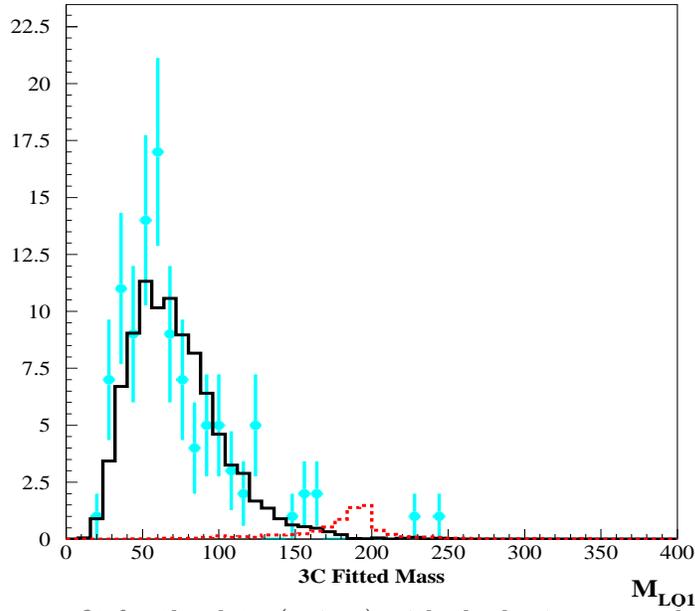,height=3.5in,width=4.0in}}
\caption{The 3C mass fit for the data (points) with the basic event selection 
shown
with background prediction (solid line historgram) 
and 200 GeV/c$^2$ mass leptoquark sample (dashed line histogram). 
Preliminary.}
\label{fig:3cdata}
\end{figure}

We have also studied the mass properties of the 101 dielectron candidate event 
sample. We used a
3C (3 constraint) fit where we require balance of the transverse energy in 
the event and we
require the masses of the two reconstructed electron-jet systems be equal.
The combination of electron with jet is chosen such that the difference in
masses is minimized before the fitting. The 3C mass fit for a 225 GeV/c$^2$
mass leptoquark sample ($\beta = 1.0$) can be seen in Fig~\ref{fig:3cmc}. 
The peak of the
distribution is about 10\% low but the mass resolution is good at about
15 GeV/c$^2$. In Fig.~\ref{fig:3cdata} is the 3C fitted mass for the data
shown as the points with errors. The background prediction is shown as the 
solid line
histogram. The data and background prediction agree well. The dashed line
histogram is the 3C fitted mass for a 200 GeV/c$^2$ leptoquark sample. We
expect about 6 events from a 200 GeV/c$^2$ mass leptoquark. By comparison we
expect about 20 events from a 160 GeV/c$^2$ mass leptoquark and 3 events
from a 225 GeV/c$^2$ mass leptoquark.
All histograms are normalized to 123 pb$^{-1}$.

\begin{figure}
\centerline{\psfig{figure=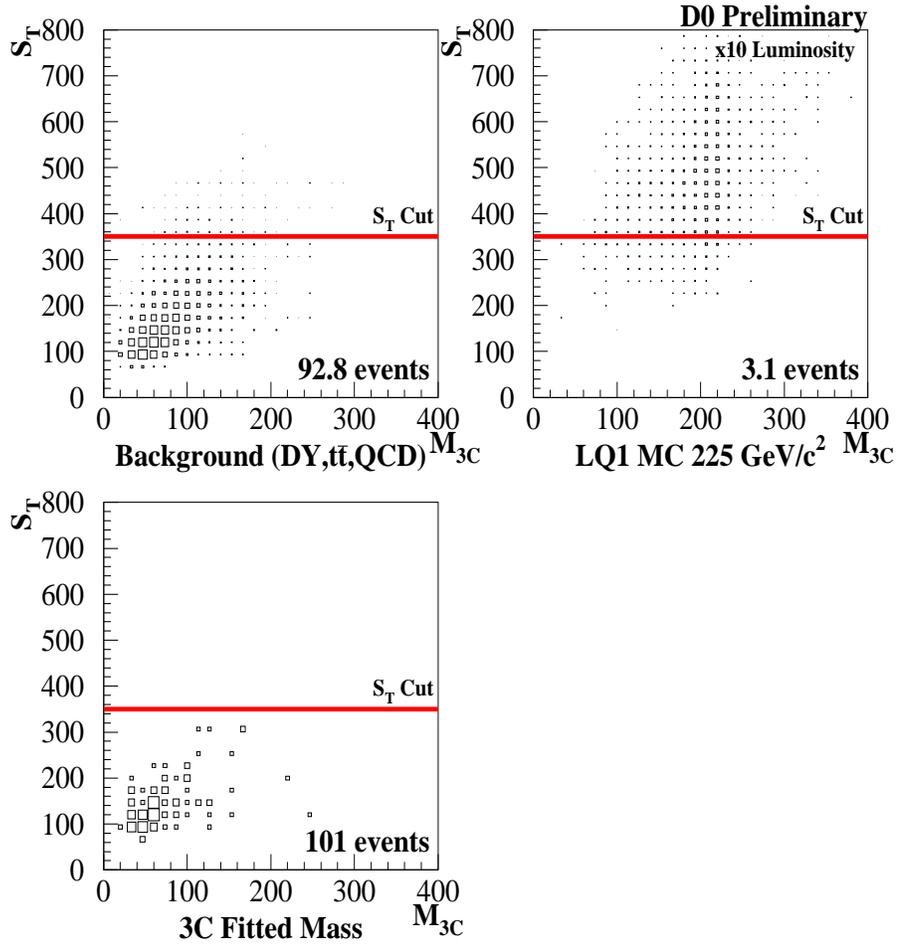,height=5.5in,width=5.0in}}
\caption{The $S_T$ vs 3C mass fit for the data with the basic event selection, 
background prediction, and 225 GeV/c$^2$ mass leptoquark sample.}
\label{fig:st3c}
\end{figure}

One might note the two events at the high end of the mass distribution in
Fig.~\ref{fig:3cdata}. These two events have values of $S_T$ that are
significantly lower than our cut of 350 GeV. This can be seen in 
Fig.~\ref{fig:st3c}. Here we plot the $S_T$ variable versus the 3C fitted
mass for the background prediction (upper left), for a 225 GeV/c$^2$ mass
leptoquark sample (upper right), and for the data (lower left). The two events
in question have $S_T$'s less than about 200 GeV. From Fig.~\ref{fig:st3c}
we see that the data and background prediction are again very similar.

\begin{figure}
\centerline{\psfig{figure=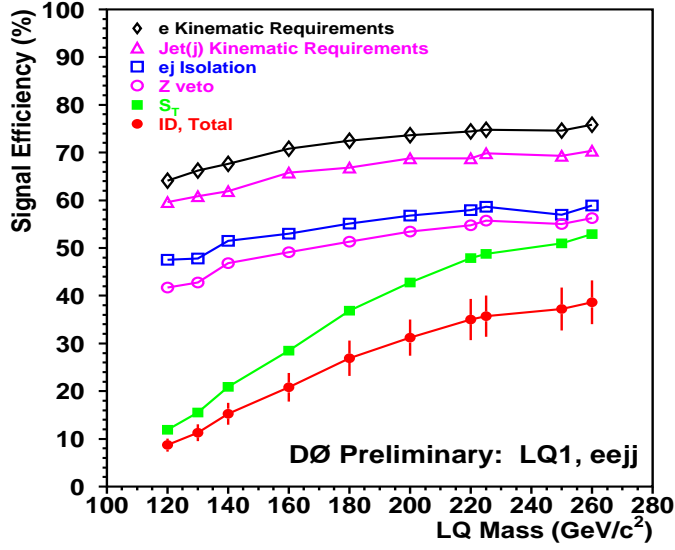,height=3.5in,width=4.0in}}
\caption{The efficiency for detection of the dielectron plus two or more jets
leptoquark signature.}
\label{fig:eff1}
\end{figure}

\begin{figure}
\centerline{\psfig{figure=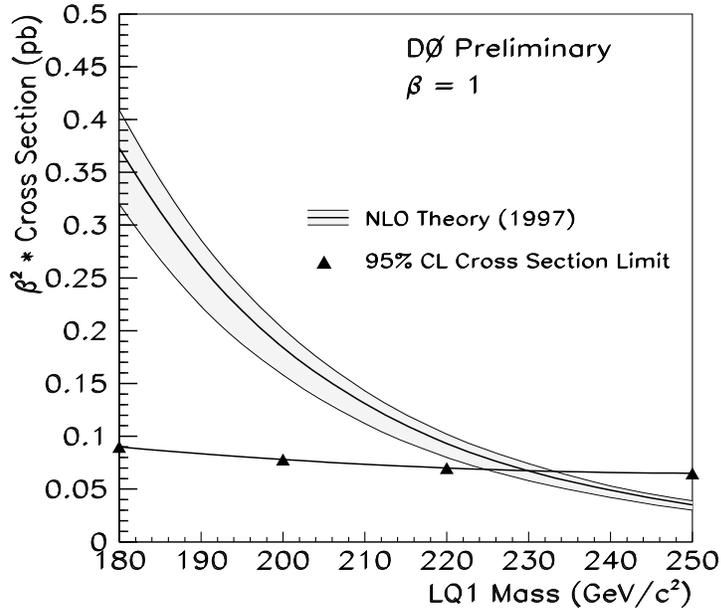,height=3.5in,width=4.0in}}
\caption{The 95\% CL limit on the production cross section times
$\beta^2$. Also shown as the band is the NLO theoretical
prediction~\protect\cite{nlo_theory}.}
\label{fig:limit1}
\end{figure}

Given that we see no dielectron leptoquark signal in our data, we can proceed 
to set a limit on the production cross section for leptoquark pairs, and by
comparing to theory we can set limits on the leptoquark mass. The
efficiency for detecting the dielectron plus two or more jet signature from
leptoquark pair production as a function of leptoquark
mass is given in Fig.~\ref{fig:eff1}. The lowest curve is the total efficiency.
The errors bars represent the total statistical plus systematic uncertainties
of about 13\%.
These uncertainties are listed here:
\begin{itemize}
\item energy scale:            \ \ \                2-5\%
\item electron identification:  \ \ \               5\%
\item acceptance:                \ \ \              5\%
\item gluon radiation:             \ \ \             7\%
\item parton distribution functions and $Q^2$: \ \ \ 7\%
\item luminosity:                             \ \ \  5\%
\item Monte Carlo statistics:            \ \ \       2\%
\end{itemize}
Given the efficiency and uncertainties, we calculate~\cite{limit_calc} 
a 95\% CL limit on the
production cross section times $\beta^2$ as a function of leptoquark mass.
This is given in Fig.~\ref{fig:limit1}. Also given is the NLO theoretical
prediction. The band represents the range in cross section prediction as
the renormalization scale changes from one half the leptoquark mass 
(upper boundary: $\mu^2 = 1/4\times M_{LQ}^2$) to twice the leptoquark
mass (lower boundary: $\mu^2 = 4 \times M_{LQ}^2$). The intersection
of our experimental limit on the cross section and the lower boundary
of the theory prediction gives a 95\% CL limit of 225 GeV/c$^2$ on the
mass of the leptoquark for $\beta = 1.0$.

We have also searched for first generation leptoquarks with the single electron
plus \MEt plus two or more jets signature. Recall that for this signature
one leptoquark decays to a electron plus quark, and the other leptoquark
decays to electron neutrino plus quark. In this case we are most sensitive
to the branching fraction $\beta = 0.5$.
We have 103.7 pb$^{-1}$ for this search. The basic event selection
is given as follows:
\begin{itemize}
\item one electron with \Et\ $>$ 25 GeV, $\mid \eta \mid < 1.2$,

\item two jets with \Et\ $>$ 25 GeV, $\mid \eta \mid < 1.0$; a third jet
with \Et\ $>$ 25 GeV is allowed in the event; additional jets with 
\Et\ $<$ 25 GeV are not used to veto events; jets are reconstructed with
cone algorithm (R = 0.7), 

\item \MEt $>$ 40 GeV, jets and \MEt must have $\Delta \phi > 0.25$ and
$\mid \pi - \Delta \phi \mid > 0.25$,

\item $S_T > 170$ GeV where $S_T = \sum{E_T^{jet}} + E_T^e$; jets must
have \Et\ $>$ than 15 GeV to be included in sum,

\item exclude events with isolated muons, since muons are not part
of the first generation leptoquark signature.

\end{itemize}
With the basic selection we find 32 candidate events. The background sources
to this signature are $W$ boson + 2 jets, top, and QCD. The background estimates
from these sources are 19.6$\pm$4.2 events from $W$ boson + 2 jets, 
9.0$\pm$2.7 events from top, and 1.1$\pm$0.4 from QCD for a total of
29.8$\pm$5.0 events. 

\begin{figure}
\centerline{\psfig{figure=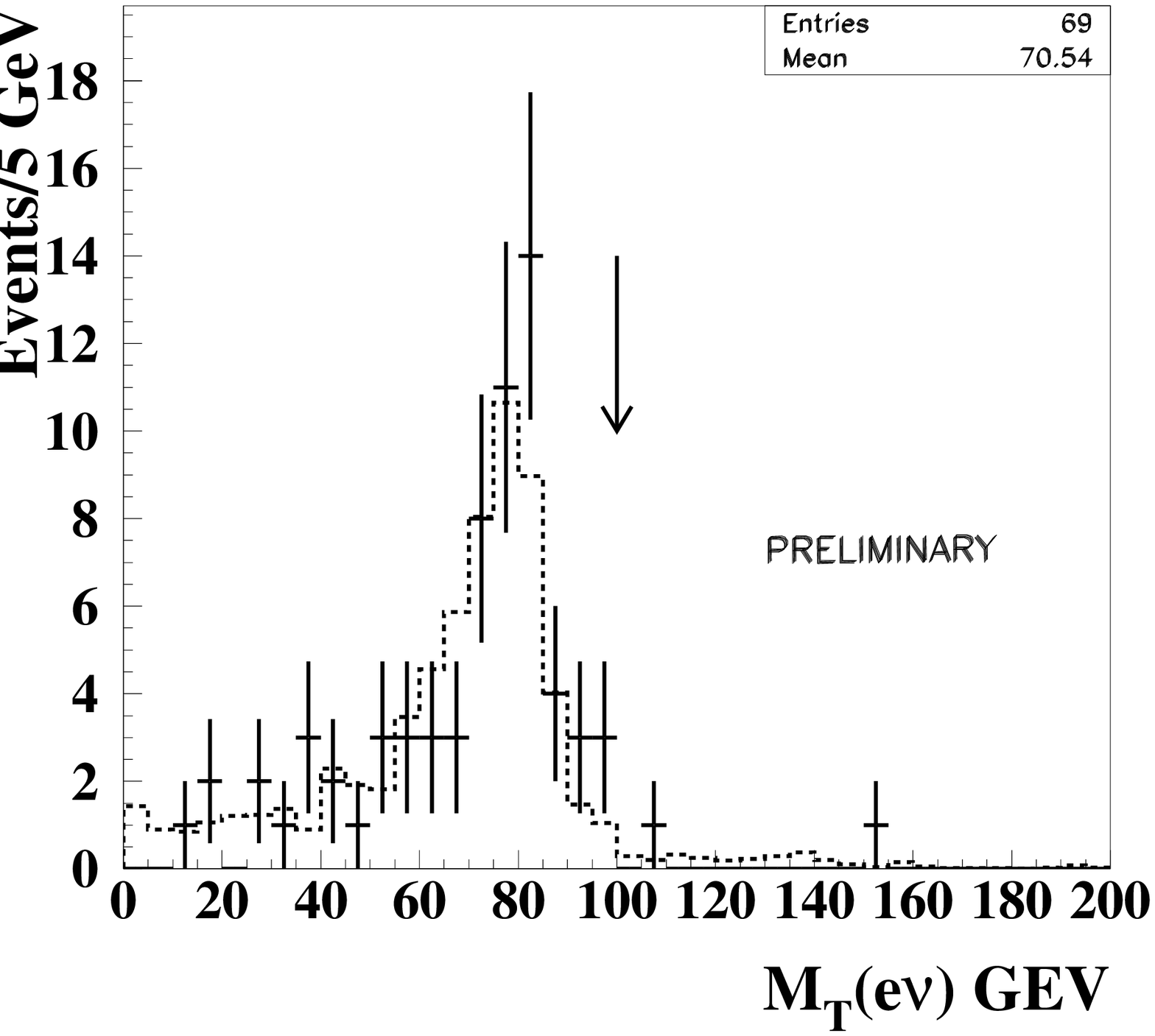,height=3.5in,width=4.0in}}
\caption{The $e-\nu$ transverse mass distribution prior to the $H_T$ cut.
The number of data events is 69. Data are the crosses, and the dashed
histogram is the background prediction. The arrow indicates the mass cut.}
\label{fig:enutm1}
\end{figure}

\begin{figure}
\centerline{\psfig{figure=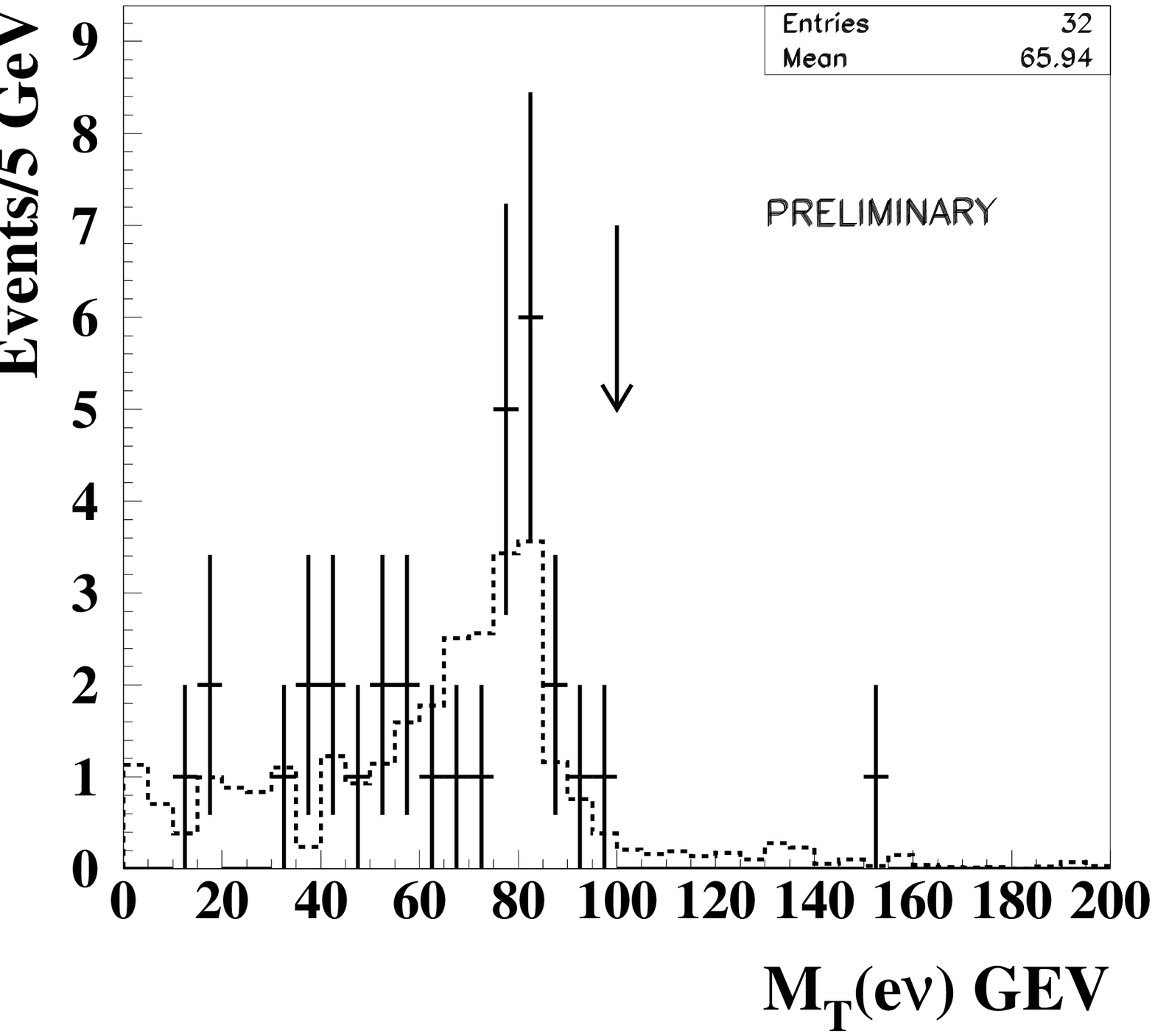,height=3.5in,width=4.0in}}
\caption{The $e-\nu$ transverse mass distribution after to the $H_T$ cut.
The number of data events is 32. Data are the crosses, and the dashed
histogram is the background prediction. The arrow indicates the mass cut.}
\label{fig:enutm2}
\end{figure}

The transverse mass of the electron and \MEt is given in Figs.~\ref{fig:enutm1}
and ~\ref{fig:enutm2}. The data is given as the points with error bars
representing the statistical uncertainties, and the background prediction is
the dashed histogram. We see very good agreement between the data
and background predictions before and after the $H_T$ cut. The arrow in the
figures represents a cut of 100 GeV on \MEt. This cut eliminates all but
one event. This event has an electron plus four jet topology and is in fact
a top candidate. This is supported by the background estimates of
0.52$\pm$0.28 events from W boson + 2jets, 1.55$\pm$0.48 events from top,
and 0.41$\pm$0.41$\pm$0.20 events from QCD for a total background prediction
of 2.5$\pm$0.6 events.

The last component of the event selection is based on a
variable, $\delta M$  defined as 
$\delta M = \mid M_{ej} - M_{LQ1}\mid / M_{LQ1}$.
$M_{LQ1}$ is the mass of the first generation leptoquark, and
$M_{ej}$ is the electron - jet invariant mass. The jet that is chosen
from the two or more jets in the event
minimizes $\delta M$. Optimizations of this variable using signal and
background Monte Carlos reveal a cut of $\delta M < 0.2$ is optimal.
We find that there are no events left in our sample for 
leptoquark masses greater than 140 GeV/c$^2$.

\begin{figure}
\centerline{\psfig{figure=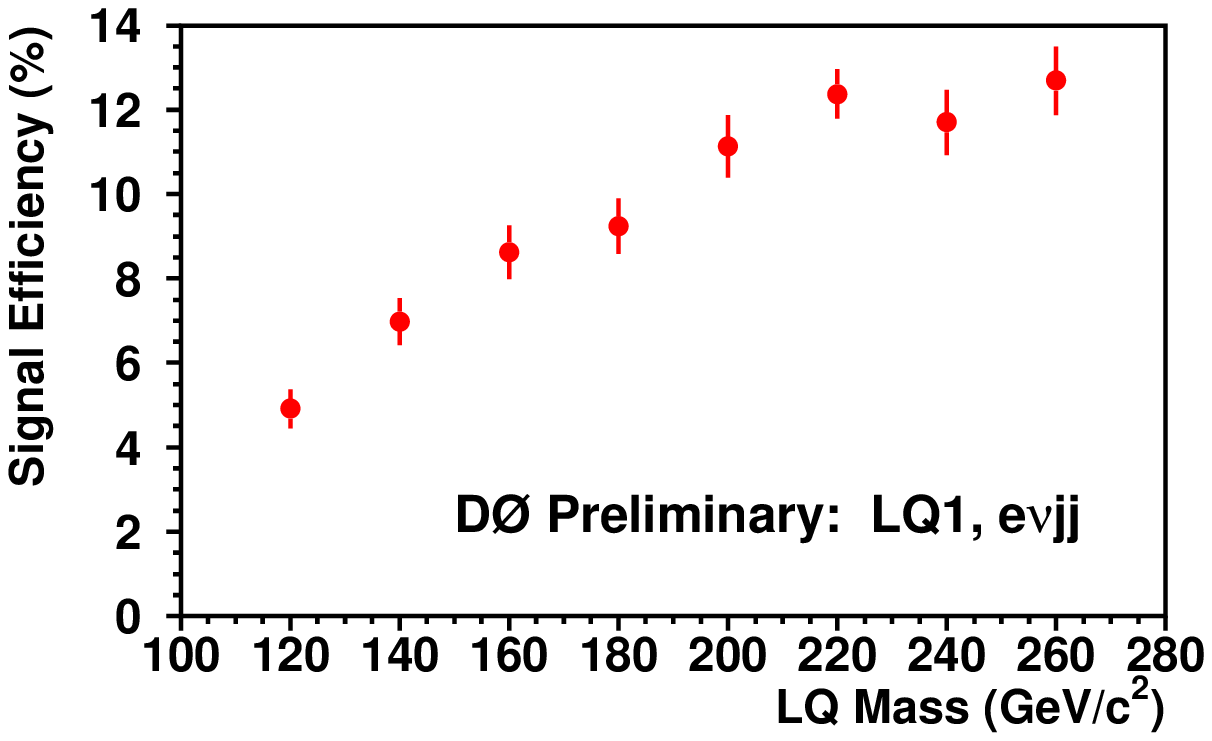,height=3.5in,width=4.0in}}
\caption{The efficiency for detection of the electron plus \MEt plus
two or more jets
leptoquark signature.}
\label{fig:eff2}
\end{figure}

\begin{figure}
\centerline{\psfig{figure=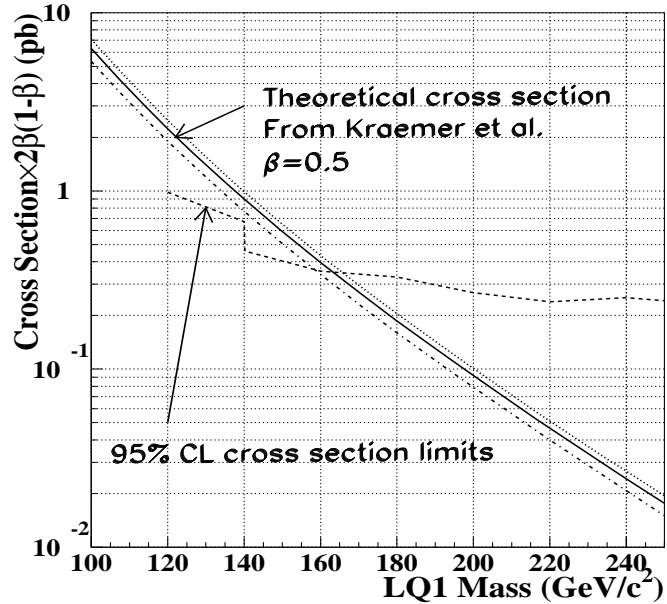,height=3.5in,width=4.0in}}
\caption{The 95\% CL limit on the production cross section times
$2\beta(1-\beta)$ for first generation leptoquarks. Also shown is the NLO 
theoretical
prediction.}
\label{fig:limit2}
\end{figure}

With
the signal efficiencies shown in Fig.~\ref{fig:eff2}, we calculate
the 95\% CL limit on the production cross section times $2\beta(1-\beta)$.
This is given in Fig.~\ref{fig:limit2} as function of leptoquark mass. 
The theoretical prediction
is the NLO production cross section times 
$2\beta(1-\beta)$ for $\beta = 0.5$.
Our 95\% CL limit drops at 140 GeV/c$^2$ because for masses greater than
140 GeV/c$^2$ our
one remaining event does not survive the $\delta M$ cut. 
The intersection of our limit
with the lower theoretical prediction ($\mu^2 = 4 \times M_{LQ}^2$) gives
a mass limit on first generation leptoquarks of 158 GeV/c$^2$ for
$\beta = 0.5$.

\section{Second Generation Leptoquarks}

\begin{figure}
\centerline{\psfig{figure=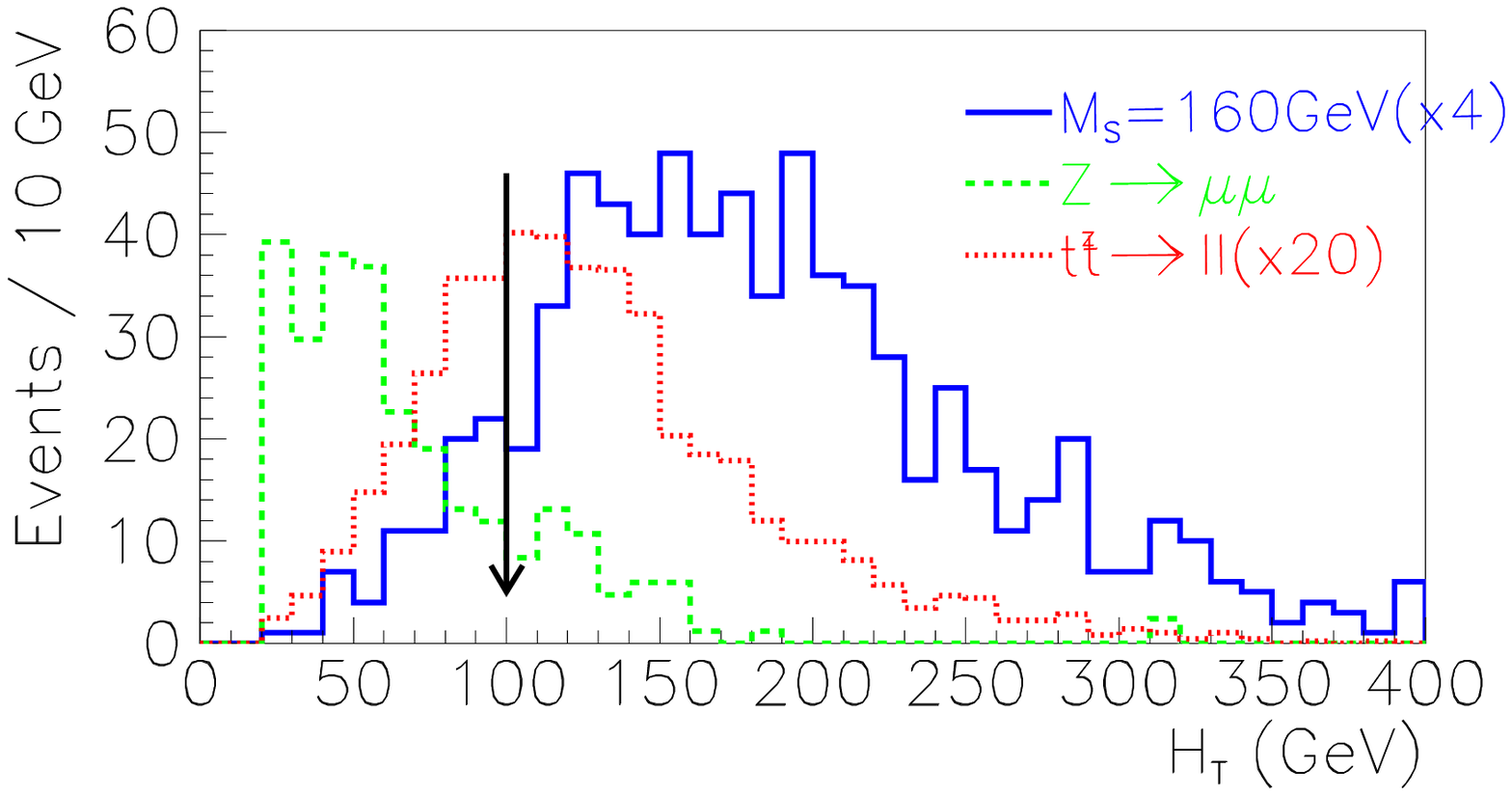,height=4.5in,width=4.0in}}
\vspace{-2in}
\caption{ The $H_T$ distribution for top and Drell-Yan backgrounds and
a 160 GeV/c$^2$ mass leptoquark signal sample. Our cut is indicated
by the arrow.
}
\label{fig:ht}
\end{figure}

The search for second generation leptoquarks involves signatures of
muons and jets. The search described here is for two isolated muons plus
two or more jets. This search uses 94.4 pb$^{-1}$ of data.
The basic event selection is the same as the standard \D0 dimuon top
quark selection~\cite{top} which because no direct \MEt cut is used
also does well for the leptoquark
dimuon event selection. The major backgrounds to second generation
leptoquarks are Drell-Yan and top. The event selection is listed here:
\begin{itemize}

\item two isolated ($R(\mu,jet)>0.5$) muons with \Et\ $>$ 15 GeV and
$\mid \eta \mid < 1.0$,

\item two jets with \Et\ $>$ 20 GeV and $\mid \eta \mid < 2.5$,

\item $\Delta \phi(\mu_1,\mu_2) < 160^o$ if $\mid \eta_{\mu_1} + \eta_{\mu_2}
\mid < 0.5$,

\item  the dimuon invariant mass, $M_{\mu\mu}$, greater than 10 GeV,

\item $H_T(jets) > 100$ GeV,

\item Z-kinematic fit probability, $\wp(\chi^2)$, less than 1\%.

\end{itemize}
Here $H_T$ is the sum of the \Et\ of jets for jets with \Et\ $>$ 15 GeV.
The
back-to-back cut in $\phi$ on the muons is for cosmic muon rejection. 
The $H_T$ cut is intended to significantly reduce the Drell-Yan background.
This can be seen in Fig.~\ref{fig:ht} where we show the $H_T$ distribution
for Drell-Yan, top, and leptoquark (M = 160 GeV/c$^2$) Monte Carlos.
We see that the $H_T$ cut of 100 GeV rejects a large portion of the
Drell-Yan ($Z\rightarrow \mu\mu$) background. 
It keeps most of the top and leptoquark
events.
With these cuts we have one event with an estimated background of
0.97$\pm$0.20 events from top and Drell-Yan.

\begin{figure}
\centerline{\psfig{figure=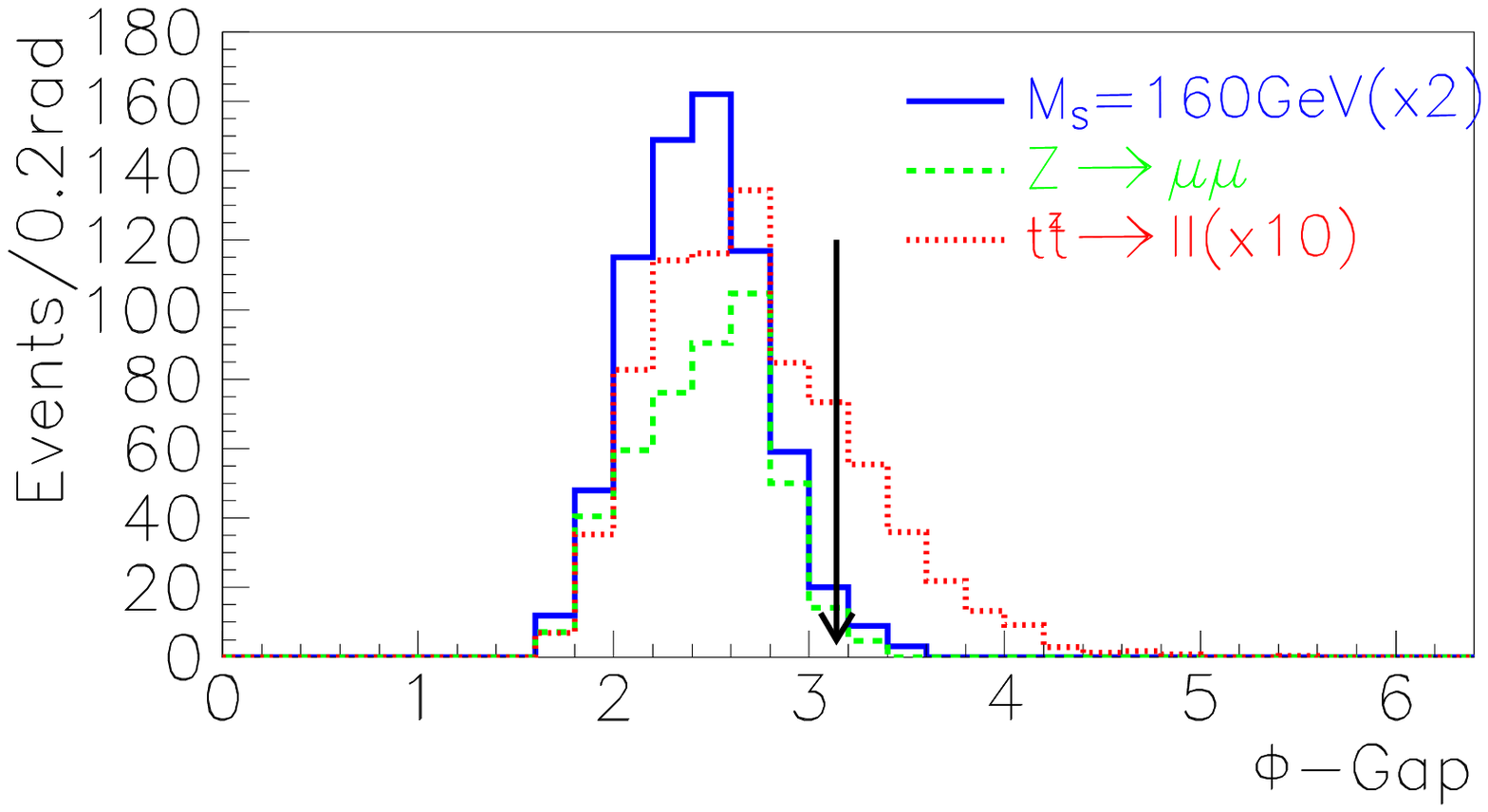,height=4.5in,width=4.0in}}
\vspace{-2in}
\caption{ The $\phi$-gap distribution for top and Drell-Yan backgrounds and
a 160 GeV/c$^2$ mass leptoquark signal sample. Our cut is indicated
by the arrow.
}
\label{fig:phigap}
\end{figure}

To reduce the top background we consider the $\phi$ distribution 
of the two muons and the two highest \Et\ jets in the candidate events.
Top pair ($t\bar{t}$) events producing the dimuon (isolated muons) signature 
contain
\MEt, the source of which is two neutrinos from the decay of the two
$W$ bosons that came from the decay of the $t\bar{t}$ pair. The dimuon 
signature from second generation 
leptoquark pair production has no \MEt from real sources like neutrinos.
It is possible that the neutrinos in the top events will concentrate
in one $\phi$ region of the detector causing the jets and muons to
concentrate on the opposite side to balance the event. This will tend to
produce a large gap in $\phi$ between two of the muons or jets. One can
imagine a pie cut up into four pieces by the two muon and two jet
directions. Using this analogy, a top event could cut out a piece of the
pie that is more than half the total pie. Leptoquark events would tend
to cut the pie into more equal pieces. The distribution of the maximum
$\phi$ gap is given in Fig.~\ref{fig:phigap} for leptoquark,
top, and Drell-Yan Monte Carlo samples.
If we require that the maximum $\phi$ gap be less than 180$^o$, we reject
a significant portion of the top background while we retain nearly all
of the leptoquark signal.

\begin{figure}
\centerline{\psfig{figure=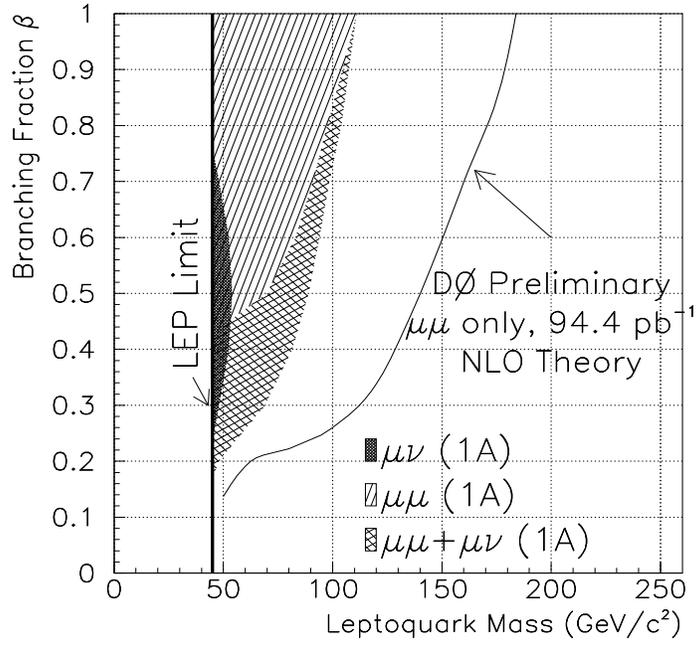,height=3.5in,width=4.0in}}
\caption{The 95\% CL limit exclusion contour for second generation leptoquarks.
Plotted is the branching fraction, $\beta$, vs leptoquark mass. We exclude
the region to the left of the curve. We use NLO theory ($\mu^2 = M_{LQ}^2$)
to determine this contour.}
\label{fig:limit3}
\end{figure}

With the $\phi$-gap cut we have no events left in our sample. The last event
is actually a $t\bar{t}$ candidate with a very large $\phi$-gap. The
total efficiencies for the signal varied from 3.8\%$\pm$0.5\% to
12.6\%$\pm$1.3\% for second generation leptoquark mass that varied from
100 GeV/c$^2$ to 260 GeV/c$^2$. The total statistical and systematic
errors are 10-15\%. We calculate the 95\% CL limit on the
production cross section time $\beta^2$ for second generation leptoquark
pairs and compare this to the NLO theory ($\mu^2 = M_{LQ}^2$). 
In Fig.~\ref{fig:limit3}
we show the preliminary 95\% CL limit exclusion contour (single line) of 
$\beta$ vs 
second generation
leptoquark mass. For $\beta = 1.0$ we have a mass limit of 184 GeV/c$^2$,
and for $\beta = 0.5$ we have a limit of 140 GeV/c$^2$. The hatched regions
are the limits from a previous analysis~\cite{d02} using only the 1992-1993 
(1A) data set.

\section{Third Generation Leptoquarks}

We've searched for signatures of third generation leptoquarks of charge 1/3.
These charge 1/3 leptoquarks decay to $b$ quark plus $\nu_\tau$. Our search
reach for third generation scalar leptoquarks is below the top mass of about
170 GeV/c$^2$, so we
assume that third generation scalar leptoquarks of charge 1/3 decay 100\%
of the time to $b$ quark plus $\nu_\tau$.

In the event selection we require \MEt $>$ 35 GeV, two jets at least one
of which has a  muon tag, and topological cuts. The untagged jets are required
to have \Et\ $>$ 25 GeV, and the tagged jets are required to have \Et\ $>$ 
10 GeV (excluding the muon \Et). We have total efficiencies of 2-5\%
for third generation leptoquark masses between 100 GeV/c$^2$ and 300 GeV/c$^2$.
The major background sources to this signature are top, $W$ and $Z$ bosons plus
two jets, and QCD multijets. For these cuts we see two events in the full
Run 1 data sample (about 20 pb$^{-1}$ for our selected trigger)
with an expected total background of 3.1$\pm$0.9 events.

\begin{figure}
\centerline{\psfig{figure=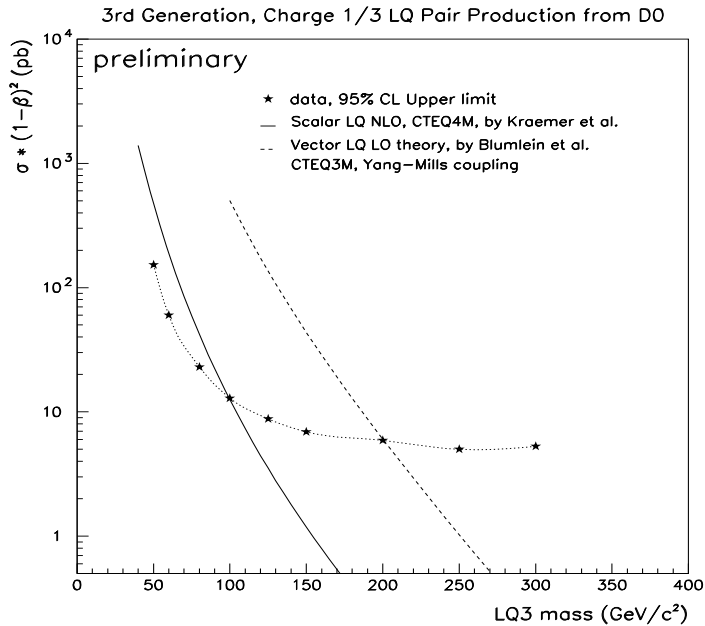,height=3.5in,width=4.0in}}
\caption{The 95\% CL limit on the production cross section for charge 1/3
third generation leptoquark.
We show here as the solid line the NLO theory ($\mu^2 = M_{LQ}^2$) for the 
scalar leptoquarks. The dashed line is LO theory for vector leptoquarks.
}
\label{fig:limit4}
\end{figure}

In Fig.~\ref{fig:limit4} we show the 95\% CL limit on the
cross section times $(1-\beta)^2$ as the stars connected by the dotted line
as a function
of third generation leptoquark mass. The solid line is the NLO theory 
($\mu^2 = M_{LQ}^2$)
for scalar leptoquarks and the dashed line is the LO theory for
vector leptoquarks with Yang-Mills coupling. From this plot we see that we
set a limit on the mass of scalar third generation leptoquarks (Q = 1/3) of
98 GeV/c$^2$, and we set a limit on third generation leptoquarks of
201 GeV/c$^2$ ($\beta = 0$).

\section{Conclusion}

We have searched for three generations of leptoquarks with diagonal couplings 
to leptons and quarks. 
We have found no evidence of a leptoquark signature
in the \D0 Run I data. A summary of the preliminary mass limits for the three
generation of leptoquarks is  given in table~\ref{tab:sum}.

For the future, we expect that our mass reach for lower $\beta$ in the second 
generation
search to improve when we add the search for the single muon plus \MEt
plus jets signature to this analysis. We expect to greatly improve our
search reach in both the first and second generation leptoquark searches
at low $\beta$ when we have added our \MEt plus jets searches to these analyses.
Finally, we will also have limits for vector leptoquarks.

\begin{table}
\caption{ Brief history of published 95\% CL mass limits on scalar leptoquarks
from hadron colliders.}
\label{tab:one}
\begin{tabular}{c c c c}
Experiment & Signature & $\beta$ & 95\% CL mass limit (GeV/c$^2$)  \\
CDF\cite{cdf1} & $eejj$ & 1.0 & 113 \\
CDF\cite{cdf1} & $eejj$ & 0.5 & 80 \\
\D0\cite{d01} & $eejj, e\nu jj$ & 1.0 & 130 \\
\D0\cite{d01} & $eejj, e\nu jj$ & 0.5 & 116 \\
HERA (H1)\cite{h1} & $ej$ &  1.0 & 275\footnote{This limit is sensitive to 
the leptoquark - lepton - quark coupling; limits from HERA assume that
$\lambda = \alpha_{\mbox{em}}$.} \\
CDF\cite{cdf2} & $\mu\mu jj$ & 1.0 & 131 \\
CDF\cite{cdf2} & $\mu\mu jj$ & 0.5 & 96 \\
\D0\cite{d02} & $\mu\mu jj, \mu\nu jj$ & 1.0 & 119 \\
\D0\cite{d02} & $\mu\mu jj, \mu\nu jj$ & 0.5 & 97 \\
CDF\cite{cdf3} & $\tau\tau jj$ & Q=4/3,2/3\footnote{ For this third generation
leptoquark limit the Top quarks is not relevant, so $\beta$ is not relevant.}
& 99 \\
\end{tabular}
\end{table}

\begin{table}
\caption{ Brief history of 95\% CL mass limits on scalar leptoquarks
from hadron colliders recently presented at conferences.}
\label{tab:two}
\begin{tabular}{c c c c}
Experiment & Signature & $\beta$ & 95\% CL mass limit (GeV/c$^2$)  \\
\D0\cite{Moriond} & $eejj, e\nu jj$ &1.0 & 175 \\
\D0\cite{Moriond} & $eejj, e\nu jj$ &0.5 & 147 \\
\D0\cite{Moriond} & $eejj, e\nu jj$ &0.0 & 81 \\
CDF\cite{vanderbilt} & $eejj$ & 1.0    & 210 (NLO)\footnote{This limit from 
CDF uses Next Leading Order (NLO) theory to set limits on the leptoquark 
mass.}\\
\D0\cite{Moriond} & $\mu\mu jj, \mu\nu jj$ & 1.0 & 167 \\
CDF\cite{Moriond} & $\mu\mu jj$ & 1.0 & 197 (NLO) \\
\D0\cite{Moriond} & $b\bar{b}\nu_{\tau}\nu_{\tau}$ & Q = 1/3 & 80 \\
CDF\cite{Moriond} & $\tau\tau jj$ & Q=4/3,2/3 & 110 (NLO) \\
\end{tabular}
\end{table}

\begin{table}
\caption{Summary of scalar leptoquark mass limits.}
\label{tab:sum}
\begin{tabular}{c c c c}
Generation & $\beta$ & 95\% CL limit (GeV/c$^2$) & comment \\
\hline
1rst & 1.0 & 225 & NLO ($\mu^2 = 4M_{LQ}^2$) \\
1rst & 0.5 & 195 & NLO ($\mu^2 = 4M_{LQ}^2$) \\
2nd & 1.0 & 184 & NLO ($\mu^2 = M_{LQ}^2$) \\
2nd & 0.5 & 140 & NLO ($\mu^2 = M_{LQ}^2$) \\
3rd & - & 98 & Q = 1/3, NLO ($\mu^2 = M_{LQ}^2$) \\
\end{tabular}
\end{table}

\end{document}